\documentclass[a4paper,10pt]{article}
\usepackage{amsthm, amssymb,amsmath}
\newcommand{\state}[1]{|#1\rangle}
\newcommand{\bra}[1]{\langle#1}

\newcommand{\cre}[2]{(a_{#2}^\dagger)^#1}
\newcommand{\des}[2]{(a_{#2})^#1}

\title{Renormalization group evaluation of exponents in family name distributions}
\author{Andrea De Luca, Paolo Rossi}

\begin{document}

\maketitle

\begin{abstract}
According to many phenomenological and theoretical studies the distribution of family name frequencies in a population can be asymptotically described by a power law.
We show that the Galton-Watson process corresponding to the dynamics of a growing population can be represented in Hilbert space, and its time evolution may be analyzed by renormalization group techniques, thus explaining the origin of the power law and establishing the connection between its exponent and the ratio between the population growth and the name production rates.
\end{abstract}
\section{Introduction}
The frequency distribution of family names in local communities, regions and whole countries has been the
object of a sustained interest by geneticists and statisticians for more than thirty years, starting from
the seminal paper by Yasuda et al. \cite{Yasuda}. For a recent review of the relevant literature we refer to
Colantonio et al. \cite{Colantonio}, while Scapoli et al. \cite{Scapoli} have recently collected and synthesized
their results on the major countries of continental Western Europe.
The main motivation for these researches resides in the deep analogy existing between surname distributions
and the frequency of neutral alleles in a population: both distributions are generated by an evolutionary
branching process subject to mutation and migration but not conditioned by natural selection.
In particular it has been observed that the dynamics of family names, in countries with an European family
name system, mimics that of the Y chromosome \cite{Sykes}.
Models for such processes have been advanced in the genetic and statistical literature, starting from the
Karlin-McGregor \cite{Karlin} statistical theory of neutral mutations. A significant theoretical evolution
occurred in particular after Lasker's empirical observation \cite{Fox} that a power law could offer a good fit of
the observed surname distributions. As a consequence Panaretos \cite{Panaretos} suggested the use of the Yule-Simon
distribution, while Consul \cite{Consul} proposed to employ the Geeta distribution with motivations coming from
a branching process modelization.
Evolutionary processes have attracted also the attention of physicists, who have found that neutral evolution
might be a ground for application of many techniques proper of statistical mechanics \cite{Derrida1} \cite{Serva}
\cite{Derrida2}.
In particular Miyazima et al. \cite{Miyazima}, studying family name distributions in Japanese towns, found the
systematic emergence of scaling laws, and further theoretical studies \cite{Zanette} \cite{Manrubia} justified the
appearance of power laws of the Yule-Simon type in the case of growing populations with non vanishing probability
for mutations.
A different explanation was offered by Reed and Hughes \cite{Reed} who considered a branching process with mutation
and migration and found that the asymptotic form of the distributions should follow a power law.
The most recent and comprehensive result is due to the Korean group of Baek et al. \cite{Baek} \cite{Kim}, who wrote down
a master equation for the frequency distribution of family names and its time evolution in the presence of birth,
death, mutation and migration, and found the possibility of different power laws with exponents depending on the
mutation and migration parameters.
In the present paper we reconsider the models of family name evolution in the context of a Hilbert space
representation of branching processes, and show that distributions characterized by an asymptotic power law
behaviour can be obtained as solutions of recursive equations which would correspond to the renormalization
group equations of an (equivalent) physical system.
In Sec. \ref{models} we introduce and motivate our models.  In Sec.\ref{immigration} we discuss the simpler case of a system characterized by pure immigration without
mutations. Finally in Sec.\ref{mutation} we discuss the case with mutation. In Appendix \ref{appendice} we represent the Galton-Watson branching process in
a Hilbert space.

\section{The models}
\label{models}
In the following sections we shall introduce two models, that take care of two different ways of generating new family names in a population:
immigration from abroad and mutation occurring after reproduction.  The importance of the appearance of new family names was pointed out in Refs. \cite{Zanette, Manrubia, Baek}.
The analogy of the recursive equations we shall obtain with those typically derived by a renormalization-group approach to a physical system will allow us
to evaluate the asymptotic behavior of the family name distribution N(k), where N is the number of family names represented by exactly k individuals.
Obviously in a typical real situation both immigration and mutation contribute to the dynamics of the family name distribution. But in our models we shall
first focus on a population in which only immigration occurs, and then on one in which only mutation occurs. This simplification is justified by the fact that
in an exponentially growing population (an approximation usually called Malthusian law) the effect of immigration can be neglected in comparison to mutation,
at least in order to study the asymptotic behavior. However, in peculiar historical conditions, mutations can be heavily depleted and as a consequence the study
of a society where name change is only due to immigration retains its value. Since we are interested in the family name distribution we can limit our attention
to the male individuals, which is consistent with the legislation on names present in most real societies. In the following we shall use the term "individual" referring
just to males. Moreover we shall suppose that the evolution of the population can be described by the Galton-Watson model. This means we shall consider:
\begin{itemize}
 \item time as discrete, moving from one generation to the next;
 \item the system as completely markovian;
 \item each individual as independent of all others.
\end{itemize}
 The last hypothesis may be considered a very strong restriction if applied to a biological system, since, for example, the exhaustion of resources induces a collective behaviour, limiting the growing rate. But we can consider this hypothesis to be valid in the context of exponential grow of a population. It is useful to fix some definitions in the use of the Galton-Watson process. We set:
\begin{equation}
\label{def_probgalt}
 p_n = \mbox{probability for an individual to have }n\mbox{ sons}
\end{equation}
It is straightforward to introduce the generating function of the Galton-Watson process:
\begin{equation}
 \label{def_galtgen}
 f(z) = \sum_{n = 1}^\infty p_n z^n
\end{equation}
Our hypothesis of growing population forces us to take $p_n$ such that the mean number of sons is greater than one:
$$ \sum_{n=1}^\infty np_n = f'(1) \equiv m > 1 $$
We will exclude the trivial case: $p_n = \delta_{1n}$.
We omit the explicit derivation of the recursive equations, which can be found with details in Appendix \ref{appendice}. However, their meaning will be somehow intuitive.

\section{Immigration}
\label{immigration}
We want to analyze a population whose members increase in number by the Galton-Watson mechanism and furthermore a group of individuals comes from outside. Each son inherits his family name from his father, while the new individuals coming from outside bring new family names. We are interested in the asymptotic behaviour of $N(k,t)$, which corresponds to the number of family names represented by $k$ individuals at time $t$. The values $N(k,0) = N_0 (k)$ are assigned as initial conditions of the problem, with:
\begin{equation}
 \label{initialcond}
 \sum_{k = 1}^\infty N_0(k) = S_0 <\infty \qquad \sum_{k=1}^\infty k N_0(k) = N_0 < \infty
\end{equation}
where $S_0$ is the initial number of family names and $N_0$ is the initial number of individuals. We introduce the generating function:
$$ n_t(z) = \sum_{k=0}^\infty N(k,t) z^k$$
Now we suppose that the individuals from outside come always distributed in the same manner: $\theta(k)$ is the number of new family names represented by $k$ individuals among them. We suppose the number of individuals $\theta_0$ and the number of new family names $G_0$ to be finite:
\begin{equation}
 \label{boundnewpeople}
\begin{array}{lll}
 \theta_0 &=& \sum_k \theta(k)\\
 G_0 &=& \sum_k k \theta(k)
\end{array}
\end{equation}
As before we introduce the generating function:
$$ \theta(z) = \sum_k \theta(k) z^k$$
We can obtain a recursive equation for $n_t(z)$ involving $\theta(z)$. The explicit derivation is given in Appendix \ref{appendice}:
\begin{equation}
 \label{eq_recursimmigr}
n_{t+1}(z) = n_t(f(z)) + \theta(z)
\end{equation}
A formal solution is given by:
$$ n_t (z) = n_0\left(f_t(z)\right) + \sum_{k=0}^{t-1} \theta\left(f_{k}(z)\right) $$
where $f_{k}(z)$ indicates the function $f(z)$ iterated $k$-times.
From this expression it is easy to compute the mean number of individuals $N_t$ and the mean number of family names $S_t$ at time $t$:
\begin{multline}
\label{totalnum}
 N_t \equiv n'_t(1)= n_0'(1)[f'(1)]^t  \sum_{k=0}^{t-1} \theta'(1) [f'(1)]^k = N_0 m^t + G_0 \sum_{k=0}^{t-1} m^k =\\
                   = \left(N_0 + \frac{G_0}{m-1}\right)m^t - \frac{G_0}{m-1}
\end{multline}
\begin{equation}
\label{totalsurname}
S_t \equiv n_t(1) = S_0 + t\theta_0 
\end{equation}
We are interested in the limit $t\to \infty$ and in the asymptotic behaviour: $k \gg 1$. In order to achieve this goal, we notice that Eq.(\ref{eq_recursimmigr}) is formally analogous to the equations coming from the renormalization group approach, linking the system at two different degrees of magnification. Therefore the system can be studied by using this analogy with the corresponding physical system. 
More explicitly, suppose $\Phi_n(T)$ is the free energy of a hierarchical model, at scale $n$ and temperature $T$. With standard renormalization group method, we can obtain the recursive equation linking two different scales (see \cite{derrida5}):
\begin{equation}
\label{eq_ricorsF}
 \Phi_{n+1}(T) = g(T) + \frac{1}{\mu} \Phi_n(\phi(T))
\end{equation}
where $g(T)$ is a regular function that comes up after summing the degree of freedom of the smaller scale and $\phi(T)$ is the RG flow. Then near the critical point, for large $n$:
\begin{equation}
\label{eq_exprg}
\Phi(T) \simeq (T-T_c)^\alpha \qquad  \alpha = \frac{\ln \mu}{\ln \phi'(T_c)}
\end{equation}
Eq.(\ref{eq_recursimmigr}) is formally analogous to Eq.(\ref{eq_ricorsF}) and in our case the role of the flow is carried out by the Galton-Watson generating function $f(z)$ and so the phases and the critical points correspond to the fixed points of $f(z)$:
\begin{equation}
 \label{eq_fixedgalt}
 f(z) = z
\end{equation}
From the fact that $f(z)$ is convex and $f'(1)>1$, we find that Eq.(\ref{eq_fixedgalt}) has three solutions: $q, 1 , \infty$\footnote{more precisely these are the possible outcomes of: $\lim_{n \to \infty} f_n(z_0)$ for different values of $z_0$.}
. From the Galton-Watson theory we know that $q \in [0,1)$ is the extinction probability. Moreover it is easy to see that $f'(q) < 1$. In fact if it was $f'(q)\geq1$ one would have by convexity:
$$ f(1) > f(q) + f'(q)(1-q) \geq 1 $$
So we have that $q, \infty$ are attractive, while $1$ is a repulsive fixed point which separates the two stable phases. We get a critical behaviour near $1$:
$$ n(z) \equiv \lim_{t\to \infty} n_t (z) \simeq (1-z)^{\alpha} $$
One can see that in this case we have that for $t \gg 1$:
\begin{equation}
 \label{eq_asymplink}
 N(k,t) \simeq k^{-1+\alpha}
\end{equation}
To compute $\alpha$, we take $\mu = 1$, $T_c = 1$, $m \equiv f'(1) = \phi(T_c)$ in Eq.(\ref{eq_exprg}) and we notice we are in an atypical situation in which $\alpha = 0$. It means that the function is diverging more slowly than any power and it is easy to see that it is logarithmic. In fact using Eq.(\ref{totalnum}) and (\ref{totalsurname}):
\begin{multline}
\label{eq_limiteimmigr}
A \equiv \lim_{z \to 1} n'(z)m^{-\frac{n(z)}{\theta_0}} = \lim_{z \to 1}\lim_{t\to\infty} n'_t(z)m^{-\frac{n_t(z)}{\theta_0}} = \\ =
\lim_{t\to \infty} \lim_{z \to 1}n'_t(z)m^{-\frac{n_t(z)}{\theta_0}} = \left(N_0+\frac{G_0}{m-1}\right)m^{-\frac{S_0}{\theta_0}}
\end{multline}
So we get near $1$
$$ n'(z) \simeq \left(N_0+\frac{G_0}{m-1}\right)m^{\frac{n(z)-S_0}{\theta_0}} = A e^{b n(z)}$$
where we set $b = \frac{\log m}{\theta_0}$. It can be solved\footnote{the arbitrary constant can be fixed by imposing the solution diverges in $1$} giving
$$ n(z) \simeq -\frac{1}{b}(\log(Ab) + \log(1-z))$$
which ensures us the logarithmic divergence and implies for large $k$:
$$N(k) = \lim_{t\to \infty} N(k,t) = \frac{C}{k}\left(1+o(1)\right) $$
So for immigrations we find a power-law behaviour with exponent $-1$. Notice that this behaviour is completely independent of the initial condition and of the distribution of the immigrating family names at each generation.
\section{Mutation}
\label{mutation}
The context is analogous to the previous one but we do no longer have immigration. We use again the initial condition in Eq.(\ref{initialcond}). Now, each son has a certain probability $\rho$ that his family name mutates into a new one, different from his father's. We suppose that $\rho$ does not depend on the family and we neglect the case in which two or more sons take the same new family name. This means the Galton-Watson contribution is modified since only a part proportional to $1-\rho$ of the offspring holds the same family name and the remaining part is added to the families of size $1$. This implies the equation:
\begin{equation}
\label{eq_mutorig}
  n_{t+1} (z) =  n_t \left(f\left(z^{1-\rho}\right)\right)+ \rho m n'_t(1) z
\end{equation}
where we used the fact that $n'_t(1)$ equals the total number of individuals at generation $t$. Observe that mutations do not contribute to the total number of individuals and so:
$$ n_t'(1) = N_0 m^t $$
as it can be shown directly via Eq.(\ref{eq_mutorig}). The recursive equation can now be solved, at least formally. Defining $r(z) = f\left(z^{1-\rho}\right)$ and indicating by $r_k(z)$ the function $r(z)$ iterated $k$-times, we get the solution:
$$ n_t (z) = n_0(r_t(z))+\rho N_0\sum_{n=0}^{t-1} m^{t-n}r_n(z) > \rho N_0 m^t r_0(z) $$
The last inequality shows that no limit in $t$ can exist. However we can obtain a limit for the function:
$$ \eta_t(z) \equiv n_t(z)m^{-t} $$
Since for large $t$: $n_t(1) \propto m^t$, as one can check by putting $z=1$ in Eq.(\ref{eq_mutorig}), we are basically considering the distribution normalized to the total number of families. So we can put Eq.(\ref{eq_mutorig}) in the form:
\begin{equation}
 \label{eq_recursmut}
 \eta_{t+1}(z) = \rho N_0 z + \frac{\eta_t (r(z)) }{m}
\end{equation}
which is again in the form of Eq.(\ref{eq_ricorsF}). However the flow is slightly changed with respect to the Galton-Watson generating function. We have $r'(1) = (1-\rho) m $ and we suppose $\rho$ small enough for $1$ to be a repulsive fixed point for the flow. In this case we must have a critical behaviour near $1$, whose exponent can be evaluated using Eq.(\ref{eq_exprg}):
$$ \eta(z) = \lim_{t \to \infty} \eta_t (z) \simeq (1-z)^\alpha$$
where the exponent can be obtained using Eq.(\ref{eq_exprg}):
$$ \alpha = \frac{\ln(m)}{\ln(r'(1))} = \frac{\ln(m)}{\ln(m)+\ln(1-\rho)} $$
Using Eq.(\ref{eq_asymplink}) we get the exponent of the family name power-law distribution:
$$\gamma \equiv \alpha + 1 = 2 - \frac{\ln(1-\rho)}{\ln(m)+\ln(1-\rho)} \simeq 2 - \frac{\rho}{\ln(m)}$$
where we considered $\rho$ very small as it is true in the real situations (see \cite{Baek}). Again the behaviour is completely independent of the initial condition and shows the typical features of a scale-free system.
\section{Conclusion}
In this paper we represented the Galton-Watson process as a quantum evolution defining the Hilbert space and the time evolution operator corresponding to the Galton-Watson probabilities. In this way we obtained two recursive equations for two possible models with different family name production mechanism: immigrations and mutations. The structure of the branching allowed us to interpret these equations as the ones that connect different scales of a physical system and, in particular, the asymptotic behaviour corresponds to the power law emerging near the critical point. The exponents are consistent with those evaluated in \cite{Baek} with a master equation approach: $N(k)$ goes as $k^{-1}$ for a society where name change is only due to immigration and, approximately, as $k^{-2}$ for a society where family name mutation occurs. Our method shows the robustness of this results, which are independent of the offspring distribution. Possible extensions of the model remain to be investigated and will be the object of future studies.
\section{Acknowledgments}
A.D.L. thanks the MIUR grant ``\textit{Fisica Statistica dei Sistemi Fortemente Correlati all'Equilibrio e Fuori Equilibrio: Risultati Esatti e Metodi di Teoria dei Campi'' -
2007JHLPEZ}, for partially supporting this work.
\appendix
\newpage
\section{The Galton-Watson process in an Hilbert space}
\label{appendice}
The structure of branching process that characterizes the Galton-Watson allows us to consider the reproduction governed by chance as a decay process whose interaction is given by an hamiltonian, which as we will see, is not hermitian. We first introduce the creation and destruction operators at each time with the usual commutation rules:
\begin{eqnarray}
[a_k, a_h] & = & 0 \\ \ [a_k^\dagger, a_h^\dagger] & = & 0 \\ \ [ a_k, a^\dagger_h] & = & \delta_{kh}
\end{eqnarray}
where, respectively, $a_k^\dagger$ creates and $a_k$ destroys an individual at time $k$. The Hilbert space is obtained in the usual way, acting on the vacuum Fock state with polynomials in $a_t^\dagger$ for all possible values of $t$. A basis for the space is then given by the following set:
\begin{equation}
\label{basis}
  \state{n,t} = (a^\dagger_t)^n \state{0} \qquad \bra{n,t}| = \bra{0}|\des {n}{t}
\end{equation}
Then, at each time $t$, the state of the system, which is determined by the probability $b_k(t)$ that exactly $k$ individuals are present, can be written:
$$\state{\phi(t)} = \sum_k b_k(t) \state{n,t}$$
It must be possible to connect the dynamics to the parameters $p_n$ introduced in Eq.(\ref{def_probgalt}) and so to the generating function $f(z)$ of Eq.(\ref{def_galtgen}). This can be done setting the hamiltonian as:
\begin{equation}
\label{def_hamil}
H(t) = f(a_{t+1}^\dagger) a_t
\end{equation}
we can write the time-evolution operator:
\begin{equation}
 \label{def_onestepevol}
 U(t) \equiv \exp(H(t))
\end{equation}
which is the one-time-step evolution operator: it evolves the states at time $t$ to time $t+1$\footnote{
    It should be observed that the correct expression for $U(t)$ should be:
    $$ U(t) = P_t e^{H(t)} $$
    where $P_t$ destroys all the states at time $t$:
    $ P_t \state 0 = \state 0$, $P_t (a^\dagger_t)^n \state 0 = 0 $ and 
    $\forall h \neq t \quad [P_t, a^\dagger_h] = 0$.
    In this way we eliminate all the parts of the states that do not evolve to time $t+1$. E.g.:
$$e^{H(t)}\state{2,t} =  \left(f\left(a^\dag_{t+1}\right)a_t + \frac{f\left(a^\dag_{t+1}\right)^2 (a_t)^2}{2}\right)(a^\dag_t)^2\state 0 = \left(2f\left(a^\dag_{t+1}\right)a^\dag_t+f\left(a_{t+1}^\dag\right)^2\right)\state 0$$
and the operator $P_t$ then eliminates the first term in the parenthesis giving the correct result.
    }, giving the correct probabilities according to the Galton-Watson process.
In fact, one can easily check that:
\begin{equation}
 \label{Uaction}
U(t)\state{1,t} = U(t) a^\dagger_t \state{0} = f(a^\dagger_{t+1})\state 0 = \sum_n p_n (a^\dag_t)^n \state{0}
\end{equation}
And in general, by linearity we know that given a state $\state{\phi(t)}$ with a particular probability distribution we get the state at time $t+1$ correctly evolved. Starting from a state $\state{\phi_0}$ at time $0$ we can obtain the state at time $T$ by:
\begin{equation}
\label{allstepevol}
  \state{\phi_T} = \mathcal U(T) \state{\phi_0} \equiv U(T-1)U(T-2)\cdots U(0)\state{\phi_0}
\end{equation}
and this equation defines the full time-evolution operator.
We see now how to derive Eq.(\ref{eq_recursimmigr}). We want to write the equation of evolution for:
$$ \state{n(t)} = \sum_{k=0}^\infty N(k,t)\state{k,t}$$
With the notation of section \ref{immigration}, we define the state:
$$ \state{\theta(t)} = \sum_{k=0}^\infty \theta(k) \state{k,t}$$
In the absence of immigration the evolution would be simply given by Eq.(\ref{allstepevol}). But here at each step, the number of family names represented by $k$ individuals grows due to the individuals coming from outside. So we get the equation:
\begin{equation}
 \label{quantumevol}
 \state{n(t+1)} = U(t)\state{n(t)}+\state{\theta(t)}
\end{equation}
Now we use the map $\mathcal W$ from the Hilbert space to $C^\infty[0,1]$ defined on the basis of Eq.(\ref{basis}) as:
\begin{equation}
 \label{def_reprhilb}
 \state{n,t} = \cre{n}{t} \state0  \quad \xrightarrow{\mathcal{W}} \quad z^n_t 
\end{equation}
And in general: $\mathcal{W}(\state{\phi(t)}) = \phi(z_t) \in C^\infty[0,1]$.
The action of an operator becomes an integral transformation. For $U(t)$ we have a simple kernel of the form: $U(t) \to U(z_t,z_{t+1}) = \delta(z_t-f(z_{t+1}))$ as can be deduced from Eq.(\ref{Uaction}). Then:
$$ \phi_{t+1}(z_{t+1}) = \mathcal W (U(t)\state{\phi(t)})= \int U(z_t,z_t+1) \phi(z_t,t) dz  = \phi_t(f(z_{t+1}))$$
where $\phi_t(z) = \mathcal W (\state{\phi(t)})$. Acting on the Eq.(\ref{quantumevol}) with the map $\mathcal{W}$ we get Eq.(\ref{eq_recursimmigr}).
\bibliography{letter}
\bibliographystyle{ieeetr}

\end{document}